\documentclass[aps,prl,preprint,groupedaddress,showpacs]{revtex4-1}

\usepackage{graphicx}
\usepackage{mathrsfs} 
\usepackage{amsmath} 

\def\bvec#1{\mbox{\boldmath $#1$}}

\begin{document}


\title{Entanglement Entropies of Non-Equilibrium Finite-Spin Systems}


\author{Koichi Nakagawa}
\affiliation{Hoshi University, Tokyo 142-8501, Japan}



\begin{abstract}
For the purpose of clarifying a new approach to understanding quantum entanglement using thermofield dynamics (TFD), entanglement entropies of non-equilibrium finite-spin systems are examined for both traditional and extended cases. The extended entanglement entropy, $\hat{S}$, is derived, and it is found that the conditions for the maximum entangled state can be obtained through this approach. The capacity of the TFD-based method to distinguish between states in quantum systems is confirmed.
\end{abstract}

\pacs{03.65.Ud, 11.10.-z, 05.70.Ln, 05.30.-d}

\maketitle

Recently, a new approach to understanding quantum entanglement using thermofield dynamics (TFD) \cite{Fano,Prigogine,Takahashi} has been proposed in Ref.~\cite{Hashizume}. In this new treatment of quantum entanglement with TFD, extended density matrices have been formulated on the double Hilbert space (ordinary and tilde Hilbert spaces), and the entanglement states show a quantum-mechanically complicated behavior. The new TFD-based method allows the entanglement states to be easily understood, because the states in the TFD tilde space play the role of tracers of the initial states. In the new analysis, a general formulation of the extended density matrices has been constructed and applied to some simple cases. Consequently, it has been found that the intrinsic quantum entanglement can be distinguished from the thermal fluctuations included in the definition of the ordinary quantum entanglement at finite temperatures. Based on the analysis presented in Ref.~\cite{Hashizume}, it has been argued that the new TFD-based method is applicable not only to equilibrium states but also to non-equilibrium states. However, analysis of the entanglement entropies of non-equilibrium systems was not conducted in Ref.~\cite{Hashizume} and, therefore, examination of the entanglement entropies of non-equilibrium systems with the use of TFD is of current interest. In the present communication, therefore, the ``extended'' entanglement entropies of non-equilibrium spin systems are intensively investigated in both the dissipative and non-dissipative cases, based upon a TFD algorithm.

Let us consider the $S=1/2$ spin system described by the Hamiltonian
\begin{align}
\mathcal{H}=-J\bvec{S}_{\text{A}}\cdot
\bvec{S}_{\text{B}}, \label{eq:1}
\end{align}
incorporating the spin operators, $\bvec{S}_{\text{A}}=(S_{\text{A}}^x, S_{\text{A}}^y, S_{\text{A}}^z)$ and $\bvec{S}_{\text{B}}=(S_{\text{B}}^x, S_{\text{B}}^y, S_{\text{B}}^z)$, of the subsystems, A and B, respectively. The state, $|s\rangle$, of the total system is then denoted by the direct product, $|s\rangle=|s_{\text{A}},s_{\text{B}}\rangle=|s_{\text{A}}\rangle |s_{\text{B}}\rangle$. Using the base, $\left\{ |++\rangle, |+-\rangle, |-+\rangle, |--\rangle \right\} $, the matrix form of the Hamiltonian \eqref{eq:1} is then expressed as \begin{align}
\mathcal{H}&=\sum _{s_{\text{A}},s_{\text{B}},s'_{\text{A}},s'_{\text{B}}}
h_{s_{\text{A}},s_{\text{B}},s'_{\text{A}},s'_{\text{B}}}|s_{\text{A}},s_{\text{B}}\rangle \langle
s'_{\text{A}},s'_{\text{B}}|\nonumber \\ &=-\frac{J}{4}\left( |++\rangle\langle++|+|--\rangle\langle--| \right) \nonumber \\
&\hspace{3mm}+\frac{J}{4}\left( |+-\rangle\langle+-| +|-+\rangle\langle-+|\right) \nonumber \\ &\hspace{3mm}-\frac{J}{2}\left(
|+-\rangle\langle-+| +|-+\rangle\langle+-|\right) .\label{eq:1a}
\end{align}
For the equilibrium states in terms of the Hamiltonian expressed in Eq.~\eqref{eq:1a} , the ordinary density matrix, $\rho _{\rm eq}$, of this system can be obtained as
\begin{align}
\rho_{\text{eq}}&:=\frac{e^{-\beta \mathcal{H}}}{Z(\beta)}\nonumber \\ &=\frac{e^{-\beta J /4}}{Z(\beta)} \left( e^{\beta J /2}\left(
|++\rangle\langle++| +|--\rangle\langle--| \right) \right. \nonumber \\ &\hspace{12mm}+\cosh \frac{\beta J }{2}\left(
|+-\rangle\langle+-|+|-+\rangle\langle-+| \right) \nonumber \\ &\left. \hspace{12mm}+\sinh \frac{\beta J }{2}\left(
|-+\rangle\langle+-|+|+-\rangle\langle-+| \right) \right) ,\label{eq:1b}
\end{align}
where $\beta $ is the inverse temperature and the partition function, $Z(\beta)$, is defined as
\begin{equation}
Z(\beta):={\text{Tr}}e^{-\beta\mathcal{H}}=2e^{-\beta J /4}\left( e^{\beta
J /2}+\cosh \frac{\beta J}{2} \right).\label{eq:1c}
\end{equation}

Let us turn our attention to a non-equilibrium system with dissipation, which is described by the Hamiltonian of Eq.~\eqref{eq:1a}. The time dependence of the ordinary density matrix, $\rho (t)$, of this system is given by the dissipative von Neumann equation \cite{Suzukibook,Suzuki1,Suzuki2}, where
\begin{equation}
i\hbar\frac{\partial}{\partial t}\rho(t)=[\mathcal{H},\rho(t)]-\epsilon \left( \rho(t)-\rho_{\text{eq}}
\right),\label{eq:2}
\end{equation}
with $\epsilon $ being a dissipation parameter. The solution of Eq.~\eqref{eq:2} is expressed as
\begin{equation}
\rho(t)=e^{-\epsilon t}U^{\dagger}(t)\rho_0 U(t)+(1-e^{-\epsilon t})\rho_{\text{eq}},\label{eq:3}
\end{equation}
for any initial density matrix, $\rho_0$, where the unitary operator, $U(t)$, denotes
\begin{align}
U(t):=e^{i\mathcal{H}t/\hbar}&=e^{i\omega t/4} \left( e^{-i\omega t/2}\left( |++\rangle\langle++| +|--\rangle\langle--| \right) \right.
\nonumber \\ &\hspace{12mm}+\cos \frac{\omega t }{2}\left( |+-\rangle\langle+-|+|-+\rangle\langle-+| \right) \nonumber \\ &\left.
\hspace{12mm}-i\sin \frac{\omega t }{2}\left( |-+\rangle\langle+-|+|+-\rangle\langle-+| \right) \right), \label{eq:7}
\end{align}
 and $\omega:= J/\hbar$. Because the explicit expression of $\rho(t)$ in Eq.~\eqref{eq:3} is complicated for any initial condition, hereafter, let us confine ourselves to the initial condition, $\rho_0=|+-\rangle\langle+-|$. Inserting Eqs.~\eqref{eq:1b} and \eqref{eq:7}, along with the initial condition, into Eq.~\eqref{eq:3}, we then obtain
\begin{align}
\rho(t)=\frac{e^{-\epsilon t }}{2}&\left( \frac{2 \left(e^{\epsilon t}-1\right)}{3+e^{-\beta \omega \hbar }}\left( |++\rangle\langle++|+|--\rangle\langle--| \right) \right.\nonumber \\ &+\frac{
\cos \omega t +e^{\epsilon t }+e^{\beta \omega \hbar } \left(3 \cos \omega t +e^{\epsilon t }+2\right)}{ 1+3 e^{\beta \omega \hbar
}}|+-\rangle\langle+-|\nonumber \\ &+\frac{ -\cos \omega t +e^{\epsilon t }+e^{\beta \omega \hbar } \left(-3 \cos \omega t +e^{\epsilon t}+2\right)}{ 1+3 e^{\beta \omega \hbar }}|-+\rangle\langle-+|\nonumber \\ &+\left(\frac{\left(e^{\epsilon t }-1 \right)
\left(-1+e^{\beta \omega \hbar }\right)}{1+3 e^{\beta \omega \hbar }}-i \sin \omega t \right)|+-\rangle\langle-+|\nonumber \\ &\left.+\left(\frac{\left(e^{\epsilon t }-1 \right) \left(-1+e^{\beta \omega \hbar }\right)}{1+3 e^{\beta \omega \hbar }}+i \sin \omega t\right)|-+\rangle\langle+-| \right) .\label{eq:08}
\end{align}

\begin{figure}[hbpt]
\begin{center}
\unitlength 1mm
\begin{picture}(130,80)
\put(0,43){
\scalebox{0.6}{
\put(0,0){\includegraphics{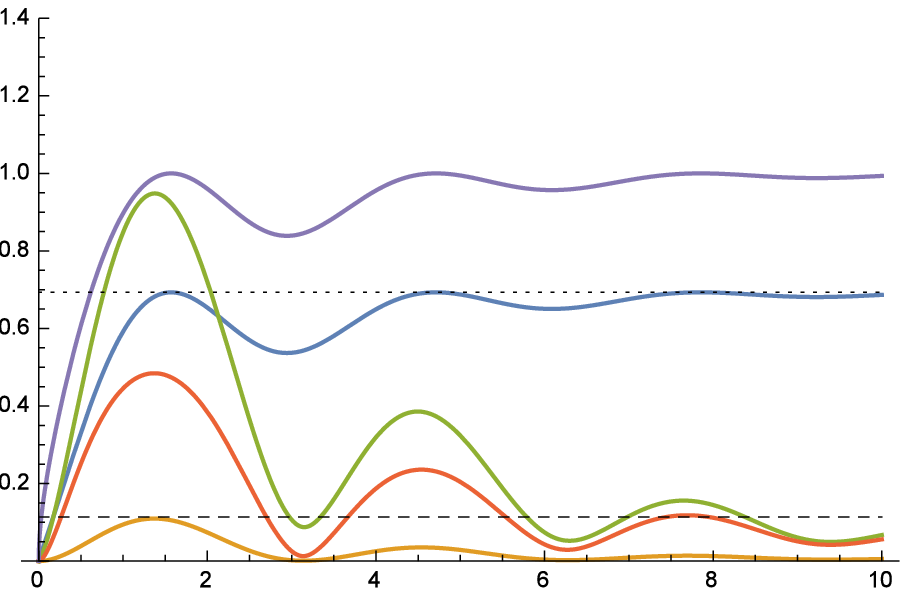}}
\put(95,0){$\omega t$}
\put(14,36){$\hat{S}$}
\put(16,31.5){$S$}
\put(16,9){$b_{\text{qe}}$}
\put(16,23){$\hat{S}_{\text{qe}}$}
\put(-1.5,30){\scalebox{0.6}{\boldmath$\log 2$}}
\put(-1.5,7){\scalebox{0.6}{\boldmath$0.11$}}
\put(16,44){$C$}
\put(47,-7){\scalebox{1.5}{(a)}}
}
}
\put(70,43){
\scalebox{0.6}{
\put(0,0){\includegraphics{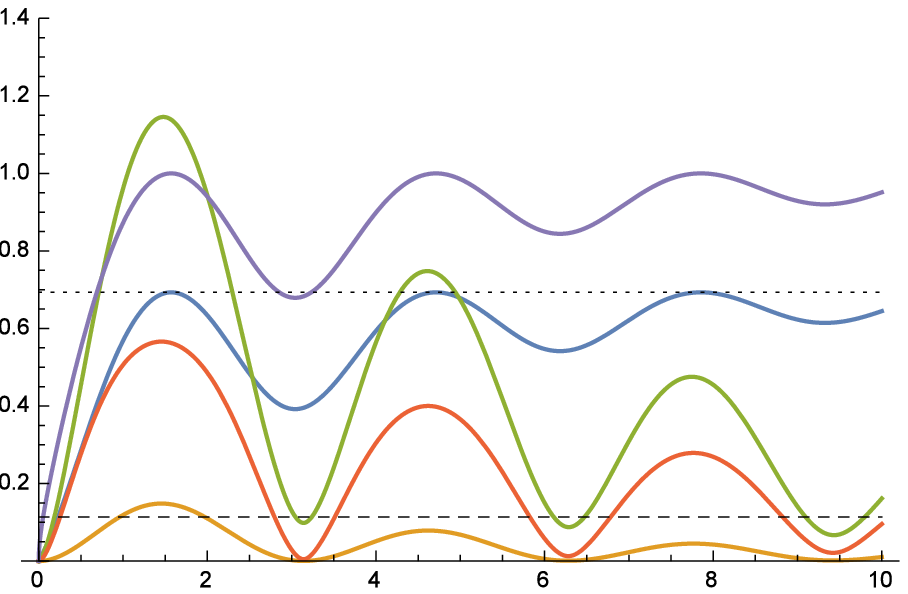}}
\put(95,0){$\omega t$}
\put(16,49){$\hat{S}$}
\put(16,31.5){$S$}
\put(16,10.5){$b_{\text{qe}}$}
\put(16,21){$\hat{S}_{\text{qe}}$}
\put(-1.5,30){\scalebox{0.6}{\boldmath$\log 2$}}
\put(-1.5,7){\scalebox{0.6}{\boldmath$0.11$}}
\put(16,39){$C$}
\put(47,-7){\scalebox{1.5}{(b)}}
}
}
\put(0,0){
\scalebox{0.6}{
\put(0,0){\includegraphics{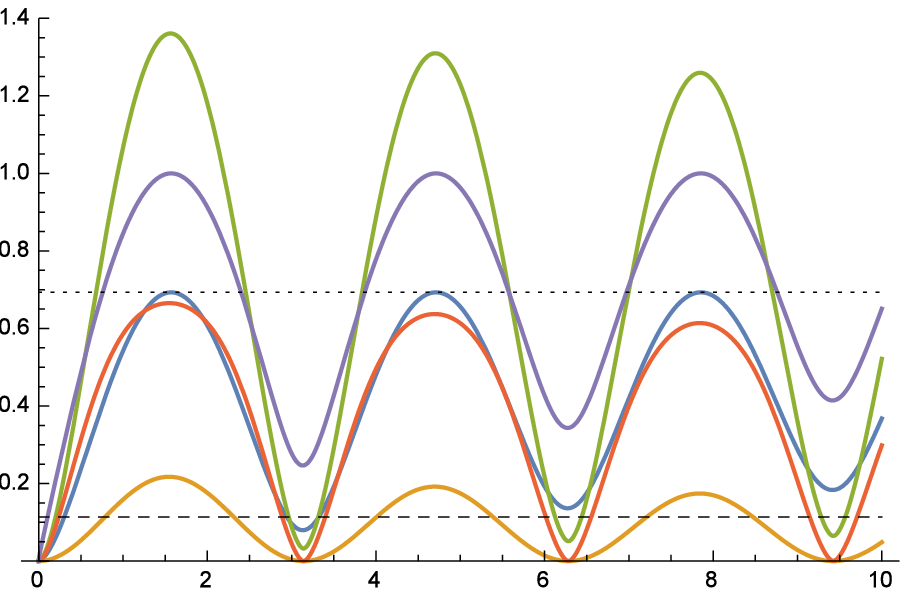}}
\put(95,0){$\omega t$}
\put(16,52){$\hat{S}$}
\put(16,31.5){$S$}
\put(16,13){$b_{\text{qe}}$}
\put(15,25){$\hat{S}_{\text{qe}}$}
\put(-1.5,30){\scalebox{0.6}{\boldmath$\log 2$}}
\put(-1.5,7){\scalebox{0.6}{\boldmath$0.11$}}
\put(15.5,39){$C$}
\put(47,-7){\scalebox{1.5}{(c)}}
}
}
\put(70,0){
\scalebox{0.6}{
\put(0,0){\includegraphics{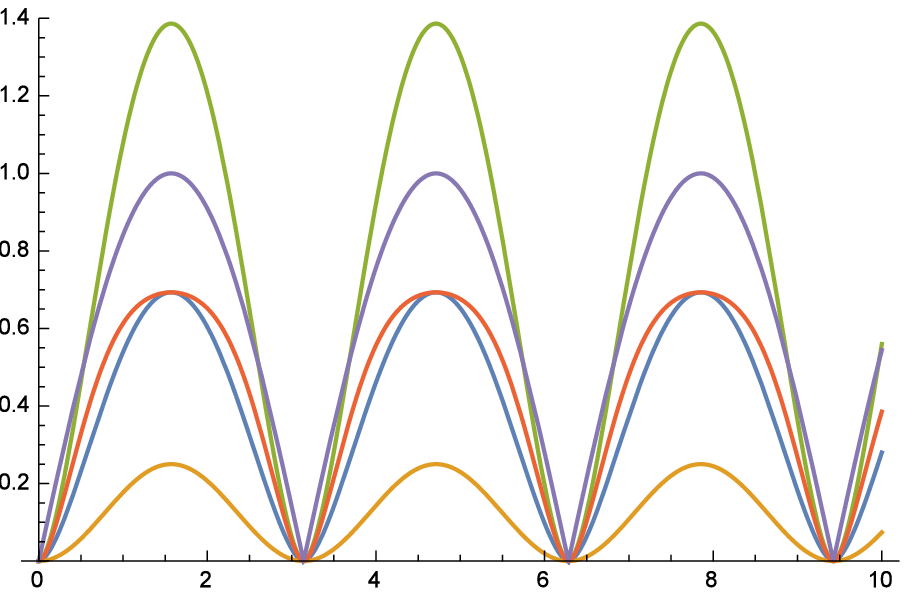}}
\put(95,0){$\omega t$}
\put(16,53){$\hat{S}$}
\put(18,24){$S$}
\put(16,14.5){$b_{\text{qe}}$}
\put(18,31){$\hat{S}_{\text{qe}}$}
\put(15.5,39){$C$}
\put(47,-7){\scalebox{1.5}{(d)}}
}
}
\end{picture}
\end{center}
\caption{Time dependence of entropies $C,~S,~\hat{S}$ and $\hat{S}_{\text{qe}}$, along with parameter $b_{\text{qe}}$, in dissipative and non-dissipative systems with scaled temperature, $T/J=0.7$. Figures (a), (b) and (c) show cases with a scaled dissipation rate of $\epsilon /\omega =0.2,~0.1$ and $0.01$, respectively. The dotted and dashed lines in Figs. (a), (b) and (c) represent the asymptotes of the $S$ and $\hat{S}$ curves, respectively. Figure (d) is the non-dissipation case ($\epsilon =0$).}\label{fig:01}
\end{figure}

The ordinary entanglement entropy, $S$, is defined by
\begin{align}
S&:=-k_{\text{B}}{\text{Tr}}_{\text{A}}\left[ \rho_{\text{A}}\log \rho_{\text{A}} \right] , \label{eq:09}
\end{align}
with $\rho_{\text{A}}:={\text{Tr}}_{\text{B}}\rho(t) $, where ${\text{Tr}}_{\text{A}}$ and ${\text{Tr}}_{\text{B}}$ represent traces over the variables of  subsystems A and B, respectively. The insertion of Eq.~\eqref{eq:08} into Eq.~\eqref{eq:09} yields
\begin{align}
S&=-k_{\text{B}}\left( \frac{1+e^{-\epsilon t}\cos \omega t}{2}\log \frac{1+e^{-\epsilon t}\cos \omega t}{2}+\frac{1-e^{-\epsilon t}\cos \omega
t}{2}\log \frac{1-e^{-\epsilon t}\cos \omega t}{2} \right) . \label{eq:11}
\end{align}
It is also possible to argue that $S$ in Eq.~\eqref{eq:11} is directly proportional to an entanglement, $E(C)$, which is a function of the ``concurrence'', $C:=\sqrt{1-e^{-2\epsilon t}\cos ^2\omega t}$ \cite{Wootters}. The time dependence of $S$ and $C$ is displayed in Fig.~\ref{fig:01} (in units of $k_{\text{B}}=1$). In the dissipative system, $S$ and $C$ converge to the constants, $k_{\text{B}}\log 2$ and $1$, respectively, at $t\to \infty $, so it is reasonable to think that $S$ and $C$ include not only the contribution of the quantum fluctuation, but also the contribution of the classical and thermal fluctuations. However, this fact is not manifest in the above expressions of $S$ and $C$.

We are now in a position to investigate the extended density matrix, $\hat{\rho }$, in the TFD double Hilbert space. Note that $\hat{\rho }$ has been defined in Ref.~\cite{Hashizume} as follows:
\begin{align}
\hat{\rho }:=|\Psi \rangle\langle\Psi |,\ |\Psi \rangle:=\rho (t)^{1/2}\sum _{s}|s,\tilde{s}\rangle=\rho (t)^{1/2}\sum _{s}|s\rangle|\tilde{s}\rangle, \label{eq:12}
\end{align}
using the ordinary density matrix, $\rho (t)$, in Eq.~\eqref{eq:3}, where $\left\{ |s\rangle \right\}$ is the orthogonal complete set in the original Hilbert space and $\left\{ |\tilde{s}\rangle \right\}$ is the same set in the tilde Hilbert space of the TFD \cite{Suzuki3,Suzuki4}. If entanglement subsystems A and B are being examined, each of the $|s\rangle$ and $|\tilde{s}\rangle$ states are represented as the direct products, $|s_{\text{A}}, s_{\text{B}}\rangle=|s_{\text{A}}\rangle | s_{\text{B}}\rangle$ and $|\tilde{s}_{\text{A}},
\tilde{s}_{\text{B}}\rangle=|\tilde{s}_{\text{A}}\rangle |\tilde{s}_{\text{B}}\rangle$, respectively.
We are then led to the renormalized extended density matrix, $\hat{\rho }_{\text{A}}$, as
\begin{align}
\hat{\rho}_{\text{A}} &:={\text{Tr}}_{\text{B}}\hat{\rho}:=\sum_{s_{\text{B}},\tilde{s}'_{\text{B}}}\langle s_{\text{B}},\tilde{s}'_{\text{B}}|\hat{\rho}|s_{\text{B}},\tilde{s}'_{\text{B}}\rangle , \nonumber \\
&=b_{\text{d1}}|+\rangle\langle+||\tilde{+}\rangle\langle\tilde{+}| +b_{\text{d2}}|-\rangle\langle-||\tilde{-}\rangle\langle\tilde{-}|
\nonumber \\ &+b_{\text{cf}}\left(
|+\rangle\langle-||\tilde{+}\rangle\langle\tilde{-}|+|-\rangle\langle+||\tilde{-}\rangle\langle\tilde{+}| \right)\nonumber \\
&+b_{\text{qe}}\left( |+\rangle\langle+||\tilde{-}\rangle\langle\tilde{-}|+|-\rangle\langle-||\tilde{+}\rangle\langle\tilde{+}|
\right), \label{eq:17}
\end{align}
where the matrix elements $b_{\text{d1}}, b_{\text{d2}}, b_{\text{cf}}$ and $b_{\text{qe}}$ are respectively obtained as analytic functions of $t, \beta, \epsilon$ and $\omega$, and correspond to the two diagonal components (d1 and d2), the classical fluctuations (cf) and the quantum entanglements (qe) of $\hat{\rho}_{\text{A}}$, respectively.
The parameter, $b_{\text{qe}}$, in Eq.~\eqref{eq:17} expresses the quantum entanglement effect.
This quantum fluctuation is essential to quantum systems, and it has been used as an order parameter of quantum systems 
\cite{Stephan,Tanaka}.
The time dependences of the parameter, $b_{\text{qe}}$, in several cases are shown in Figs.~\ref{fig:01}.
As can be seen from Eq.~\eqref{eq:17}, only the intrinsic quantum entanglement is extracted clearly in the TFD formulation.
In particular, it can be understood that the entangled state of the system emerges through a single product, such as $|+\rangle\langle +||\tilde{-}\rangle\langle \tilde{-}|$, in $\hat{\rho}_{\text{A}}$.

The ``extended'' entanglement entropy is defined as
\begin{align}
\hat{S}&:=-k_{\text{B}}\mathrm{Tr}_{\text{A}}\left[ \hat{\rho}_{\text{A}}\log \hat{\rho}_{\text{A}} \right] ,
\label{eq:22}
\end{align}
using the renormalized $\hat{\rho }_{\text{A}}$ in Eq.~\eqref{eq:17} \cite{Hashizume}. The insertion of Eq.~\eqref{eq:17} into Eq.~\eqref{eq:22} and subsequent simplification eventually yield
\begin{equation}
\hat{S}=\hat{S}_{\rm cl}+\hat{S}_{\rm qe},
\label{eq:23}
\end{equation}
where
\begin{align}
\hat{S}_{\rm cl}&:=-k_{\text{B}}\left( \sqrt{4b_{\text{cf}}^2+\left(b_{\text{d1}}-b_{\text{d2}}\right)^2} ~\mathrm{arccoth}\,
\frac{b_{\text{d1}}+b_{\text{d2}}}{\sqrt{4b_{\text{cf}}^2+\left(b_{\text{d1}}-b_{\text{d2}}\right)^2}}\right.\notag \\
&\hspace{3mm}\left. +\frac{b_{\text{d1}}+b_{\text{d2}}}{2}\log \left(b_{\text{d1}}b_{\text{d2}}-b_{\text{cf}}^2\right)\right),
\label{eq:23a}
\end{align}
and
\begin{align}
\hat{S}_{\rm qe}:=-2k_{\text{B}}b_{\text{qe}}\log b_{\text{qe}},
\label{eq:23b}
\end{align}
respectively.
In Eqs.~\eqref{eq:23}, \eqref{eq:23a} and \eqref{eq:23b}, the expressions of $\hat{S}$, the classical and thermal fluctuation parts, $\hat{S}_{\rm cl}$, and the quantum entanglement part, $\hat{S}_{\rm qe}$, also incorporate analytic functions of $ t, \beta, \epsilon$ and $\omega$, respectively; however, the full calculation is quite tedious.
So, we show the numerical behaviour of $C,~S,~\hat{S},~\hat{S}_{\text{qe}}$ and $b_{\text{qe}}$ for a few cases in Figs.~\ref{fig:01}(a)$-$(c) (in units of $k_{\text{B}}=1$).
As can be seen from these figures, at $t \to \infty $, $\hat{S}$ converges to the value, $0.1135\cdots$, and both $\hat{S}_{\text{qe}}$ and $b_{\text{qe}}$ vanish, respectively. As a consequence, the traditional entanglement entropy, $S$, becomes larger than the extended entanglement entropies, $\hat{S}$ and $\hat{S}_{\rm qe}$, at $t \to \infty $.
$\hat{S}_{\text{qe}}$ is then smaller than $S$ when $\epsilon $ is relatively larger. As $\epsilon $ becomes smaller,  $\hat{S}_{\text{qe}}$ becomes compatible with $S$ and, at $\epsilon =0$, $\hat{S}_{\text{qe}}\lesssim S$. These results suggest that the quantum entanglement, $\hat{S}_{\text{qe}}$, is enhanced as the dissipation becomes weaker.


For non-dissipative systems, $\hat{S}$ in Eq.~\eqref{eq:23} and $\hat{S}_{\rm qe}$ in Eq.~\eqref{eq:23b} reduce to
\begin{align}
\hat{S}&=-k_{\text{B}}\left( \cos ^4 \frac{\omega t}{2}\cdot \log \left( \cos ^4 \frac{\omega t}{2} \right) +  \sin ^4 \frac{\omega t}{2}\cdot \log \left( \sin ^4 \frac{\omega t}{2} \right)  + \frac{1}{2} \sin ^2 \omega t\cdot \log \left( \frac{\sin ^2 \omega t}{4} \right)\right),
\label{eq:24}
\end{align}
and 
\begin{align}
\hat{S}_{\rm qe}=- \frac{k_{\text{B}}}{2} \sin ^2 \omega t\cdot \log \left( \frac{\sin ^2 \omega t}{4} \right),
\label{eq:25}
\end{align}
respectively, at $\epsilon =0$. The time dependence of $\hat{S}$ and $\hat{S}_{\rm qe}$ at $\epsilon =0$ is shown in Fig.~\ref{fig:01}(d) (in units of $k_{\text{B}}=1$). It is apparent in this figure that all the curves ($C,~S,~\hat{S},~\hat{S}_{\text{qe}}$ and $b_{\text{qe}}$) showing the entanglement have the same phase, however, their amplitudes differ.
Specifically, $\hat{S}$ is larger than $S$ and $\hat{S}_{\rm qe}\approx S$ at $\epsilon =0$, a result which differs from that of Ref.~\cite{Hashizume} and which can be seen in Eqs.~\eqref{eq:24} and \eqref{eq:25}. 
It appears that a mistake was made in Ref.~\cite{Hashizume} in counting the non-zero eigenvalues of $\hat{\rho }_{\text{A}}$.



In this communication, we have examined the extended entanglement entropies of non-equilibrium spin systems in both the dissipative and non-dissipative cases, based upon the TFD formulation. In the dissipative case in particular, the extended entanglement entropy is derived using the extended density matrix and is proven to separate into the classical and thermal fluctuation parts and the quantum entanglement part. These quantities are compared to the traditional entanglement entropy, the concurrence, and $b_{\text{qe}}$ in $\hat{\rho }_{\text{A}}$.
These results are summarized in Figs.~\ref{fig:01} and show that the conditions yielding the maximum entangled state can be obtained using these five quantities.

We have clearly indicated that, in the TFD-formulation, the extended quantum entanglement entropy part and the parameter $b_{\text{qe}}$ are recognized as effective quantities for measurement of the quantum entanglement.
It is apparent that the new TFD-based method enables us to clearly distinguish between the various states of quantum systems.

\bibliography{eenefss_revtex_refs}
\end{document}